\documentclass{article}

\usepackage{mathrsfs}
\usepackage{eufrak}

\newcommand{\be}{\begin{equation}}
\newcommand{\ee}{\end{equation}}
\newcommand{\ba}{\begin{eqnarray}}
\newcommand{\ea}{\end{eqnarray}}

\newcommand{\sym}{$\mathcal{N}=4$ SYM }
\newcommand{\msl}{$\mathfrak{sl}(2)$ }
\newcommand{\der}{\mathcal{D}}
\newcommand{\tr}{\mathrm{Tr}\,}

\title{On the large spin limit of \\ twist operators in \sym}

\author{Francesca Catino and Matteo Beccaria\\
  Dipartimento di Fisica, Universita' del Salento\\
  INFN Sezione di Lecce
  }

\begin{document}

\maketitle

\begin{center}
\begin{minipage}[h]{10cm}
\centerline{\bf Abstract}
\medskip
The long range Bethe Ansatz solution of the mixing problem in \sym allows to compute in a very efficient way multiloop anomalous dimensions of various composite operators. 
In the case of \msl twist operators it is important to obtain closed expressions for the anomalous dimensions in terms of the Lorentz spin.
Conjectures are available altough analytical proofs are missing beyond one-loop. In this paper, we will present a method to expand at large spin the solution of the 
long range Baxter equation in twist 2 and 3. We will also propose sum rules for special singlet states at higher twist.
\end{minipage}
\end{center}

\section{Introduction}

\sym is the maximal supersymmetric Yang-Mills theory in four dimensions.
It has become of great interest in the last decade thanks to Maldacena conjecture \cite{Mal} suggesting that  \sym is dual to type IIB superstring theory on $AdS_5\times S^5$. 
In particular, the anomalous dimensions of gauge invariant composite operators of \sym are related by duality 
to the string energy levels energy. From the classical integrability of string theory on $AdS_5\times S^5$, we infer that the dilatation operator
can be regarded as an integrable {\em hamiltonian} in the planar limit.
This  internal integrability of \sym has become clear since the seminal paper \cite{MZ} where Minahan and Zarembo showed that in the planar limit single trace operators of \sym could be interpreted in terms of (super)spin chains with energy equal to the anomalous dimension of the operator. Since these spin chains are integrable, they could be treated by means of the 
integrability machinery and in particular Bethe Ansatz \cite{Fad} or Baxter operators \cite{Baxter} methods.

In this note we restrict our consideration to the so-called \msl sector of \sym, which is an invariant subsector closed under perturbative renormalization mixing. 
It is spanned by single trace operators
\be\label{Intro:singleTraceOP}
\mathscr{O}_{n_1, \dots, n_L} = \tr \left( \der^{n_1}\,\varphi\,\der^{n_2}\,\varphi\,\cdots \der^{n_L}\,\varphi\right), \quad \quad n_1 + \cdots + n_L = N
\ee
where $\varphi$ is one of the three complex scalar fields of \sym, $\der$ is a light cone covariant derivative, $\{ n_i \}$ are non negative integers and their sum $N$ is the total spin. The number $L$ of fields $\varphi$ is called twist of the operator and it is equal to the classical dimension minus the spin. The \msl sector is very reach and interesting, in particular for certain similarities with analogous QCD operators appearing in deep inelastic scattering.

Scaling composite operators are linear combinations of operators $\mathscr{O}_{n_1, \dots, n_L}$ of \msl sector which are eigenvectors of the dilatation operator with 
eigenvalues the anomalous dimensions $\gamma(N,L,\lambda)$. $\lambda$ is the 't Hooft planar coupling
\be\label{Intro:lambda}
\lambda=\frac{g_{YM}^2\, N_c}{8\,\pi^2}.
\ee
At fixed $L$, we are interested in the perturbative expansion 
\be\label{Intro:perturbGexp}
\gamma(N,\lambda)=\sum_{n=1}^{\infty}\gamma_n(N)\,\lambda^n.
\ee
Technically, integrability permits to write down Bethe Ansatz equations that compute $\gamma$ order by order in $\lambda$ and that can be solved for each given $N$.
However it is not obvious how to find a parametric expression of $\gamma_n(N)$ as a closed function of the spin $N$, order by order in 
perturbation theory.
Nevertheless such closed formulae are important in physical applications where, for instance, one is interested in various limits and 
analytic continuations over $N$.

From this point of view, it is necessary to separate the twist 2 and 3 case from the higher twist one.
At twist 2 and 3 there are conjectured closed formulae for the ground state (with minimal anomalous dimension) up to three loops. However proofs are missing, at least beyond one loop.
At higher twist no simple closed expressions describe the ground state, whose anomalous dimension is irrational.

In this note we describe a method ($\Delta$-method) that permits to derive large spin expansion of $\gamma_n(N)$ at twist 2 and 3 without assuming any conjecture, 
directly from the long range Baxter equation.
Moreover, at higher twist we provide quite simple sum rules for anomalous dimensions of the ground and excited states, parametrically in both $N$ and the twist $L$.  

\section{The long range Baxter equation in the \msl sector}

Before we proceed any further, let's remind the structure of the Baxter equation in \msl sector.
At one loop, it reads
\be\label{BaxtEq:oneLoop}
(u+i/2)^L\,Q(u+i)+(u-i/2)^L\,Q(u-i)=t_L(u)\,Q(u)
\ee
where
\ba\label{BaxtEq:tAndq_2}
t_L(u) & = & 2\,u^L+q_{2,\,L}\,u^{l-2}+ \cdots + q_{L,\,L}\\
q_{2,\,L} & = & -(N+L/2)\,(N+L/2-1)+L/4.
\ea
We shall be interested in a polynomial solution
\be\label{BaxtEq:BaxterPolyn}
Q(u)=\prod_{j=1}^{N}(u-u_j).
\ee
Replacing (\ref{BaxtEq:BaxterPolyn}) into (\ref{BaxtEq:oneLoop}) we obtain the Bethe Ansatz equations
\be\label{BaxtEq:BAE}
\left(\frac{u_k+i/2}{u_k-i/2}\right)^L=\prod_{j=1,\,j\neq k}^N\frac{u_k-u_j-i}{u_k-u_j+i}
\ee
The one loop anomalous dimension is than 
\be\label{BaxtEq:gamma1}
\gamma_1=i\,(\log Q(u))'\bigg|_{u=-i/2}^{u=+i/2}.
\ee

At higher loop, the long range Baxter equation for \msl sector of \sym is discussed at 3 loops accuracy in \cite{MVV}.

Let's define
\ba\label{BaxtEq:modCoord}
x(u) & = & \frac{u}{2}\left(1+\sqrt{1-\frac{2\,\lambda}{u^2}}\right)\\
x_{\pm} & = & x(u\pm i/2)
\ea
and for $\sigma=\pm1$
\ba\label{BaxtEq:modFunc}
\Lambda_{\sigma}^{(n)} & = & \frac{d^n}{d\,u^n}\,\log Q(u)\bigg|_{u=\frac{i}{2}\,\sigma}\\
\Delta_{\sigma}(x) & = & x^3\,\exp \left(-\frac{\lambda}{x}\,\Lambda_{\sigma}^{(1)}-\frac{\lambda^2}{4\,x^2}(\Lambda_{\sigma}^{(2)}+x\,\Lambda_{\sigma}^{(3)})\right).
\ea
The Baxter equation reads now
\be\label{BaxtEq:threeLoop}
\Delta_+(x_+),Q(u+i)+\Delta_-(x_-)\,Q(u-i)=t_L(u)\,Q(u)
\ee
with
\be\label{BaxtEq:tLhigherLoop}
t_L(u)=2\,u^L+q_{1,\,L}\,u^{L-1}+q_{2,\,L}\,u^{L-2}+\cdots q_{L,\,L}
\ee
and 
\ba\label{BaxtEq:charges}
q_{1,\,L} = & & \lambda\,q_{1,\,L}^{(1)} + \lambda^2\,q_{1,\,L}^{(2)}, \\
q_{i,\,L} = & q_{i,\,L}^{(0)} + & \lambda\,q_{i,\,L}^{(1)} + \lambda^2\,q_{i,\,L}^{(2)}
\ea
for $i=2,\cdots L$. The equation must be solved in terms of
\be\label{BaxtEq:BaxterPolynHL}
Q(u)=Q^{(0)}(u)+\lambda\,Q^{(1)}(u)+\lambda^2\,Q^{(2)}(u)
\ee
and
\be\label{BaxtEq:AnomalDimHL}
\gamma=\gamma_1\,\lambda+\gamma_2\,\lambda^2+\gamma_3\,\lambda^3=i\,\lambda\left[(\log Q(u))'+\frac{\lambda}{4}\,(\log Q(u))'''+\frac{\lambda^2}{48}\,(\log Q(u))'''''\right]_{u=-i/2}^{u=+i/2}.
\ee

\section{Large spin expansion in twist 2 and 3}

The conjectures for $\gamma_n(N)$ that we want to check are the following ones. In twist 2, the two loop anomalous dimension
reads (even $N$) \cite{Kotikov},
\ba
\gamma_2 &=& -4\,\left[S_3(N)+S_{-3}(N)-2\,S_{-2,1}(N)+2\,S_1(N)\,(S_2(N)+S_{-2}(N))\right] = \nonumber \\ \nonumber\\
&=& -S_3\left(\frac{N}{2}\right)+8\,S_{-2,1}(N)-4\,S_1(N)\,S_2\left(\frac{N}{2}\right).
\ea
In twist 3, the two and three loop anomalous dimension reads
\ba
\gamma_2 &=& -2\,S_3-4\,S_1\,S_2, \\
\gamma_3 &=& 5\,S_5 + 6\,S_{2}\,S_{3} - 8\,S_{3, 1, 1} + 4\,S_{4, 1} - 4\,S_{2, 3} +  \\
&& + S_{1}\,(4\,S_{2}^2 + 2\,S_{4} + 8\, S_{3, 1}),\nonumber
\ea
with all harmonic sums evaluated at $N/2$~\cite{Beccaria:2007cn,Kotikov:2007cy}.
Harmonic sums are defined as usual by
\be
S_a(N) = \sum_{n=1}^N\frac{(\mbox{sign}\,a)^n}{n^{|a|}},\qquad
S_{a, \mathbf{b}} = \sum_{n=1}^N\frac{(\mbox{sign}\,a)^n}{n^{|a|}}\,S_{\mathbf{b}}(n).
\ee

As a first step in checking these conjectures one can try to recover their large spin expansion.
Here we illustrate a method that permits to achieve this goal in the twist 2 and 3 cases.
We start in the easy one loop case, where we rephrase a technique originally devised by G. Korchemsky in \cite{Kor}. The resulting procedure will be called $\Delta$-method. It works well and the desired expansion is systematically obtained in few lines of calculation, easily implemented on symbolic algebra packages.

Then we move to two and three loop cases. Here we find that $\Delta$-method fails. We propose a safe improved expansion in these cases.
Note that we shall work with even $N$.

\subsection{$\Delta$-method}

Here we present the $\Delta$-method in a very short way. For more details see \cite{LongRange}.

Let's start with one loop case and define
\be\label{Tw2e3:epsilon}
\varepsilon=\frac{1}{N}\; \textrm{(twist 2)},\quad \varepsilon=\frac{2}{N}\; \textrm{(twist 3)} 
\ee
and for $u=iz$
\be\label{Tw2e3:QeD}
Q(iz)=e^{F(z)}, \quad \Delta(z)=F(z+1)-F(z).
\ee
Using the definitions we can rewrite the Baxter equation. For instance at twist 2 we have
\be
\label{Tw2e3:BEtwist2}
\varepsilon^2\,\left(z+\frac{1}{2}\right)^2\,e^{\Delta(z)}+\varepsilon^2\,\left(z-\frac{1}{2}\right)^2\,e^{-\Delta(z-1)} = 1+\varepsilon+\varepsilon^2\,\left(
\frac{1}{2}+2\,z^2\right).
\ee
Than we put the asymptotic expansion
\be
\label{Tw2e3:delta}
\Delta(z) = -2\,\log\varepsilon + \sum_{n=0}^\infty a_n(z)\,\varepsilon^n.
\ee
in the  Baxter equation and we match the various integer powers of $\varepsilon$ to obtain the functions $a_n(z)$.
Given $\Delta(z)$, we get back to derivative of $\log Q$ that permits to calculate the one loop anomalous dimension. We than immediately recover the correct expansion in full agreement with the conjectures.

At two (or three) loops we have
\ba
F(z) & = & F^{(0)}(z)+\lambda\, F^{(1)}(z)\: \Big(\,+\lambda^2\,F^{(2)}(z)\Big)\\
\Delta(z) & = & \Delta^{(0)}(z)+\lambda\,\Delta^{(1)}(z) \: \Big(\,+\lambda^2\,\Delta^{(2)}(z)\Big)
\ea
where $\Delta^{(0)}(z)$ has just been found and for the new contribution $\Delta^{(1)}(z)$ (the same for $\Delta^{(2)}(z)$) we put
\be\label{Tw2e3:delta1}
\Delta^{(1)}(z) = \sum_{n=0}^\infty (a_n(z)\,\log\varepsilon+b_n(z))\,\varepsilon^n
\ee
in the Baxter equation and proceeding as one loop we obtains the following results for twist 2
\ba
\gamma_{2, \Delta-{\rm method}} &=& \left(\frac{2}{3} \pi ^2 \log\,\overline{\varepsilon}-6 \zeta_3\right)+\left(-8 \log\,\overline{\varepsilon}-\frac{\pi ^2}{3}\right) \varepsilon +
\nonumber\\
&& + \left(4 \log\,\overline{\varepsilon}+\frac{\pi
   ^2}{18}+6\right) \varepsilon ^2 + \left(-\frac{4 \log\,\overline{\varepsilon}}{3}-\frac{14}{3}\right) \varepsilon ^3+\nonumber\\
&& + \left(2-\frac{\pi ^2}{180}\right) \varepsilon ^4+\cdots~~~.
\ea
and for twist 3
\ba
\gamma_{2, \Delta-{\rm method}} &=& \left(\frac{2}{3} \pi ^2 \log\,\overline{\varepsilon}-2 \zeta_3\right)+\left(-4 \log\,\overline{\varepsilon}-\frac{\pi ^2}{3}\right) 
\varepsilon +\\
&& + \left(2 \log\,\overline{\varepsilon}+\nonumber + \frac{\pi
   ^2}{18}+3\right) \varepsilon ^2+ \left(-\frac{2 \log\,\overline{\varepsilon}}{3}-\frac{7}{3}\right) \varepsilon ^3+\nonumber\\
&& + \left(1-\frac{\pi ^2}{180}\right) \varepsilon ^4+\left(\frac{2
   \log\,\overline{\varepsilon}}{15}-\frac{1}{45}\right) \varepsilon ^5+\left(-\frac{1}{4}+\frac{\pi ^2}{378}\right) \varepsilon ^6+\cdots  \nonumber
\ea
\ba
\gamma_{3, \Delta-{\rm method}} &=& \left(-\frac{11}{45} \pi ^4 \log\,\varepsilon-\zeta_5+\frac{1}{3} \pi ^2 \zeta_3\right)+\left(\frac{4}{3} \pi ^2 \log\,\varepsilon-2 \zeta_3+\frac{11 \pi
   ^4}{90}\right) \varepsilon +\nonumber\\
& +& \left(-2 \log^2\,\varepsilon-\frac{2}{3} \pi ^2 \log\,\varepsilon-4 \log\,\varepsilon+\zeta_3-\frac{11 \pi ^4}{540}-\frac{5 \pi
   ^2}{6}\right) \varepsilon ^2+\cdots
\ea 
Comparing with the conjectured expansion one sees that a mismatch appears for $\gamma_{2, \Delta-{\rm method}}$ at twist 2 and $\gamma_{3, \Delta-{\rm method}}$ at twist 3.
All terms of this mismatch are not transcendental neither have logarithmic enhancement.

The failure of the $\Delta$-method is related to the fact that we have assumed an expansion for $\Delta(z)$ valid in the Baxter equation for both $\Delta(z)$ and $\Delta(1-z)$ in a neighborhood of $z=1/2$. The assumed expansion is clearly wrong.

If we proceed more rigorously we find new terms that contain powers of $\varepsilon^{4\,z}$. These new anomalous terms will appear in the $F$ expansion. One could see that, considering the new anomalous terms, the results confirms the conjectures.

\section{Higher twist}

For $L>3$ the zero momentum highest weight states of the $L$ site \msl spin chain can be divided into two subsets: singlets with non degenerate energy and paired states with degeneracy 2. We are interested to singlet states.

It is easy to see that singlets are all obtain by solving Baxter equation with the requirement that transfer matrix $t_L(u)$ has definite parity
\be
t_L(u)=(-1)^L\,t_L(u).
\ee
This sets to zero several quantum numbers in $t_L(u)$. The number of singlet states for twist $L=2\,n$ or $L=2\,n+1$ is a function of the Lorentz spin
\be
\#\quad {\rm singlets}=\left(\begin{array}{c}
\frac{N}{2}+n-1\\
n-1
\end{array}\right).
\ee
Now, we can compute the sum of the anomalous dimensions of singlet states
\be
\Sigma_L^{(s)}(N) = \sum_{k\in \ \rm singlets}\gamma_{L, k}^{(s)}(N).
\ee
It is easy to see (Appendix B in \cite{SumRules}) that this quantity is rational. Given a sequence of rational numbers describing the $N$ dependence of $\Sigma_L(N)$, it is possible to look for closed formulae by using some trial and error combinations of harmonic sums.

We have extended the calculation up to $L=13$ testing the following structural properties:
\begin{enumerate}
\item The general formula for $\Sigma_L$ up to three loops takes the form 
\be
\Sigma_L(N) = \sum_{n_1=1}^{\frac{N}{2}}\,\sum_{n_2=1}^{n_1}\cdots\sum_{n_p=1}^{n_{p-1}}\,\sigma_L(n_p),
\ee
where the number of sums is $p=n-1$ for both $L=2\,n$ and $L=2\,n+1$.

\item The internal function $\sigma_L(n_p)$ can be written as a linear combination of harmonic sums with total 
transcendentality equal to $2\,\ell-1$ where $\ell$ is the loop order $\ell = 1, 2, 3$.

\item The argument of the harmonic sums is $n_p$ for odd $L$ and $2\,n_p$ for even $L$.

\item The multi-index of the harmonic sums does involve only positive indices for odd $L$.

\item The set of multi-indices is the same for all even $L$ and fixed loop order. The same is true for odd $L$ with a different 
set of indices. 
\end{enumerate}
Looking at the $L$ dependence of the coefficients of the harmonic sums we have been able to write down the following compact expression for odd $L$
\ba
\label{eq:linearodd}
\Sigma_L(N) &=&\phantom{+} 2\,(L-1)\,S_{X, 1}\,g^2 + \\
&&  + \big[(3\,L-7)\,S_{X,3}-2\,(L-1)\,S_{X, 1, 2}-4\,(L-2)\,S_{X, 2, 1}\big]\,g^4 + \nonumber\\
&&  + \big[(20\,L-79) \,S_{X,5} - 6\,(L-1) \,S_{X,1,4} -12\,(2\,L-7) \,S_{X,4,1} +\nonumber\\
&& -2\,(8\,L-21) \,S_{X,2,3} -2\,(12\,L-37) \,S_{X,3,2}
+ 4\,(L-1) \,S_{X,1,2,2} + \nonumber\\
&& + 8\,(L-2) \,S_{X,2,1,2} + 8\,(2\,L-5) \,S_{X,2,2,1} + 8\,(L-1) \,S_{X,1,3,1} + \nonumber\\
&& + 24\,(L-3) \,S_{X,3,1,1}\big]\,g^6 + \cdots ~~.\nonumber
\ea
where
\be
S_{X, \mathbf{a}}\equiv S_{X, \mathbf{a}}\left(\frac{N}{2}\right),\qquad X = \underbrace{\{0, \cdots, 0\}}_{\frac{L-3}{2}},
\ee
and for even $L$
\ba
\label{eq:lineareven}
\Sigma_L^{(1/2)}(N) &=&\left[2\,L\,\widetilde{S}_{X,1} + 2\,(L-2)\,\widetilde{S}_{X,-1}\right]\,g^2 + \\
&&  + \big[
4 (3 L-8) \,\widetilde{S}_{X,-3}+12 (L-2) \,\widetilde{S}_{X, 3}-8 (L-2) \,\widetilde{S}_{X,-2,-1} \nonumber\\
&& -8 (L-3) \,\widetilde{S}_{X,-2,1} -4 (L-2) \,\widetilde{S}_{X,-1,-2}-4 (L-2) \,\widetilde{S}_{X,-1,2} \nonumber\\
&& -4 L \,\widetilde{S}_{X,1,-2}-4 L \,\widetilde{S}_{X,1,2}-8 (L-2) \,\widetilde{S}_{X,2,-1}-8 (L-1) \,\widetilde{S}_{X,2,1}
\big]\,g^4 + \cdots~. \nonumber
\ea
where
\be
\widetilde{S}_{X, \mathbf{a}}\equiv \widetilde{S}_{X, \mathbf{a}}\left(\frac{N}{2}\right),\qquad X = \underbrace{\{0, \cdots, 0\}}_{\frac{L-2}{2}}
\ee
An important check of the previous results is that for large $N$ all the ground and excited states are expected to scale logarithmically with $N$ with a coupling 
dependence which can be reabsorbed in the so-called physical coupling (~\cite{Korchemsky:1992xv,Belitsky:2003ys,Belitsky:2006en} 
and~\cite{Eden:2006rx,Beisert:2006ez}) 
\be
g_{{\rm ph}}^2=g^2-\zeta_2\,g^4+\frac{11\,\pi^4}{180}\,g^6+\cdots
\ee
Using some useful results for harmonic sums we find for $\Sigma_L(N)$ the same asymptotic behavior of the anomalous dimensions one by one.

It is possible to derive sum rules at arbitrary high order. The determination of the explicit formulae is a matter of computational effort. In \cite{SumRules} we give quadratic and cubic sum rules.

\section{Conclusions}

It is useful for physical applications to find closed formulae for anomalous dimensions of \msl scaling operators. 
This is non trivial beyond one loop and nowadays conjectures have been given only at twist 2 and 3. 
We have found large spin expansion of anomalous dimensions at twist 2 and 3 directly from Baxter equation and it is a strong check for the conjectured closed formulae. 
It could be also useful to demonstrate some asymptotic properties like Gribov-Lipatov reciprocity \cite{GL}.

We have also obtained new closed formulae at higher twist, not for anomalous dimensions, which are generically irrational, but for sums of anomalous dimensions.

It remains to be understood if the closed expression we found for the sum rules are just a curiosity or a manifestation of deeper property, for instance interpreted in the light of AdS/CFT.


\begin{thebibliography}{99}

\bibitem{Mal}
{\sc J.M. Maldacena},  Adv. Theor. Math. Phys. 2 {\bf 231} (1998)

\bibitem{MZ}
J. A. Minahan and K. Zarembo, JHEP {\bf 0303}, 013 (2003)

\bibitem{Fad}
L.D. Faddeev, {\em Les Houches 1995, Relativistic gravitation and gravitational radiation}, 149-219 

\bibitem{Baxter}
R.J. Baxter, ``Exactly Solved Models in Statistical Mechanics'', Academic Press, London, 1982

\bibitem{MVV}

A. Vogt, S. Moch and M. Vermaseren, Nucl. Phys. {\bf B 691}, 129 (2004)

\bibitem{Kotikov}
A.~V.~Kotikov and L.~N.~Lipatov,
  Nucl.\ Phys.\  {\bf B 661}, 19 (2003)
  [Erratum-ibid.\  {\bf B 685}, 405 (2004)]

  A.~V.~Kotikov, L.~N.~Lipatov, A.~I.~Onishchenko and V.~N.~Velizhanin,
  Phys.\ Lett.\ {\bf B 595}, 521 (2004)
  [Erratum-ibid.\  {\bf B 632}, 754 (2006)]
  
\bibitem{Beccaria:2007cn}
  M.~Beccaria,
  JHEP {\bf 0706}, 044 (2007)
  
\bibitem{Kotikov:2007cy}
  A.~V.~Kotikov, L.~N.~Lipatov, A.~Rej, M.~Staudacher and V.~N.~Velizhanin,
  arXiv:0704.3586 [hep-th].

\bibitem{Kor}
G. P. Korchemscky, Nucl. Phys. {\bf B 462} (1996) 333

\bibitem{LongRange}
M. Beccaria and F. Catino, JHEP {\bf 0801}, 067 (2008)

\bibitem{Korchemsky:1992xv}
  G.~P.~Korchemsky and G.~Marchesini, 
  Nucl.\ Phys.\  {\bf B 406}, 225 (1993)

\bibitem{Belitsky:2003ys}
  A.~V.~Belitsky, A.~S.~Gorsky and G.~P.~Korchemsky,
  Nucl.\ Phys.\  {\bf B 667}, 3 (2003)


\bibitem{Belitsky:2006en}
  A.~V.~Belitsky, A.~S.~Gorsky and G.~P.~Korchemsky,
  {\em Logarithmic scaling in gauge / string correspondence}, 
  Nucl.\ Phys.\  {\bf B 748}, 24 (2006)

\bibitem{Eden:2006rx}
  B.~Eden and M.~Staudacher,
  J.\ Stat.\ Mech.\  {\bf 0611}, P014 (2006)
  
\bibitem{Beisert:2006ez}
  N.~Beisert, B.~Eden and M.~Staudacher,
  J.\ Stat.\ Mech.\  {\bf 0701}, P021 (2007)

\bibitem{SumRules}
M. Beccaria and F. Catino, JHEP {\bf 0806}, 103 (2008)

\bibitem{GL}
M. Beccaria, Yu. L. Dokshitzer and G. Marchesini, Phys. Lett. {\bf B 652}, 194 (2007)

\end{thebibliography}
\end{document}